\documentclass[iop, superscriptaddress, preprint]{revtex4-1}
\pdfoutput=1
\usepackage{graphicx}

\usepackage{amsmath}
\usepackage{amssymb}
\usepackage{bm}
\usepackage{ulem}
\usepackage{comment}

\newcommand{\bra}[1]{\langle{#1}|}
\newcommand{\ket}[1]{|{#1}\rangle}

\newcommand{\argmin}{\mathop{\rm arg~min}\limits}

\begin{document}

\title{
Sparse identification of quantum Hamiltonian dynamics via quantum circuit learning
}

\author{Yusei Tateyama}
\email{yusei.tateyama@gmail.com}
\affiliation{Department of Complex and Intelligent Systems, Future University Hakodate, Hokkaido 041-8655, Japan }

\author{Yuzuru Kato}
\email{Corresponding author: katoyuzu@fun.ac.jp}
\affiliation{Department of Complex and Intelligent Systems, Future University Hakodate, Hokkaido 041-8655, Japan }

\date{\today}

\begin{abstract}
Sparse identification of nonlinear dynamics (SINDy) is a data-driven framework for estimating classical nonlinear dynamical systems from time-series data. In this approach, system dynamics is represented as a linear combination of a predefined set of basis functions, and the corresponding coefficients are sparsely estimated from observed time-series data. In this study, we propose sparse identification of quantum Hamiltonian dynamics (SIQHDy), a SINDy-inspired quantum circuit learning framework for estimating quantum Hamiltonian dynamics from time-series data of quantum measurement outcomes. In SIQHDy, the unitary evolution of a quantum Hamiltonian system is expressed as a product of basis quantum circuits, and the corresponding circuit parameters are estimated through sparsity-promoting optimization. We numerically demonstrate that SIQHDy accurately reconstructs the dynamics of single-, three-, and five-spin systems, and exhibits robustness to measurement noise in the three-spin case.
Furthermore, we propose an extension of SIQHDy for scenarios with limited accessible observables and evaluate its performance in identifying two-spin systems and in network-structure identification for five-spin systems.

\end{abstract}

\maketitle



\section{Introduction}

Accurate modeling of dynamical systems from time-series data remains a fundamental challenge across a wide range of scientific and engineering fields, such as nonlinear dynamics, fluid mechanics, and materials science~\cite{brunton2019data, mendez2023data, himanen2019data}. In classical dynamical systems, data-driven approaches for estimating governing equations from time-series data have attracted considerable attention. One approach to modeling dynamical systems from time-series data is to construct a best-fit model that captures the essential underlying dynamics. Notable examples include the eigensystem realization algorithm (ERA)~\cite{juang1985eigensystem}, which reconstructs reduced-order models from input-output data, and dynamic mode decomposition (DMD)~\cite{schmid2010dynamic, kutz2016dynamic}, which employs a linear regression framework to extract coherently evolving structures underlying complex nonlinear dynamics.

A more challenging task is to simultaneously identify both the structure and parameters of the governing equations directly from the observed data. Sparse identification of nonlinear dynamics (SINDy)~\cite{brunton2016discovering} addresses this problem by representing system dynamics as a linear combination of a predefined set of basis functions, with the corresponding coefficients sparsely estimated from observed data (see also the earlier related work based on compressive sensing \cite{wang2011predicting}). This sparse regression framework enables the identification of dominant model structures that capture the essential features of the system dynamics while estimating the underlying governing equations.
Several variants of the SINDy framework have also been proposed, including SINDy integrated with model predictive control for data-driven control in the low-data limit~\cite{kaiser2018sparse}, SINDy for systems with hidden variables using delay coordinates~\cite{brunton2017chaos}, and extensions to implicit dynamical systems such as implicit-SINDy and SINDy-PI~\cite{mangan2017inferring,kaheman2020sindy}. Weak SINDy, which employs an integral formulation, improves robustness to noisy measurement data~\cite{messenger2021weak}, and ensemble-SINDy enhances robustness in low-data or high-noise regimes through bootstrap aggregating~\cite{fasel2022ensemble}.
SINDy and its variants have been applied to various physical systems, including reduced-order modeling in fluid dynamics~\cite{loiseau2018constrained} and plasma physics~\cite{alves2022data}. Moreover, extensions of SINDy to stochastic dynamical systems~\cite{boninsegna2018sparse, callaham2021nonlinear} and partial differential equations~\cite{rudy2017data, messenger2021weak2} have further broadened the scope of the applicability of the framework. 

Several classical data-driven approaches for estimating underlying dynamics have also been extended to quantum system identification \cite{zhang2014quantum, zhang2015identification, goldschmidt2021bilinear, kato2022data, kato2023dynamical, kaneko2025forecasting, yin2023analyzing, shabani2011estimation}. In particular, ERA-based parameter estimation methods have been developed for identifying both quantum Hamiltonian systems~\cite{zhang2014quantum} and open quantum systems~\cite{zhang2015identification}.
Regarding the DMD framework, a bilinear DMD method has been proposed for data-driven quantum control \cite{goldschmidt2021bilinear}, and we have recently developed Hankel and kernel DMD methods for quantum system identification~\cite{kato2022data, kato2023dynamical}. In addition, the DMD method has been applied to forecasting quantum many-body dynamics~\cite{kaneko2025forecasting, yin2023analyzing}.
However, to the best of our knowledge, the SINDy framework has not yet been explicitly developed for quantum systems.

Another promising direction for data-driven estimation and prediction of quantum dynamics is the use of quantum computers. With recent progress in quantum information technology, hybrid quantum-classical algorithms have emerged as practical approaches for mitigating quantum noise by combining parameterized quantum circuits with classical optimization \cite{preskill2018quantum}. Quantum circuit learning (QCL) is a well-known example, where circuit parameters are optimized for tasks such as function approximation, classification, and quantum simulation~\cite{mitarai2018quantum}.
Hamiltonian identification methods based on hybrid quantum-classical machine learning have also been proposed~\cite{gupta2023hamiltonian, shi2022parameterized} in the context of Hamiltonian learning \cite{wiebe2014hamiltonian, wang2017experimental, zubida2021optimal, yu2023robust}. In particular, the method in Ref.~\cite{gupta2023hamiltonian} typically assumes a predefined Hamiltonian structure and employs variational quantum algorithms to estimate its parameters from observed data. 
However, in some solid-state systems, for example, the Hamiltonian structure itself may be unknown. In such cases, a more challenging and important task is to simultaneously identify both the Hamiltonian structure and its parameters directly from time-series quantum measurement data.

\begin{figure} [!t]
	\begin{center} 
        \includegraphics[width=1\hsize,clip]
        {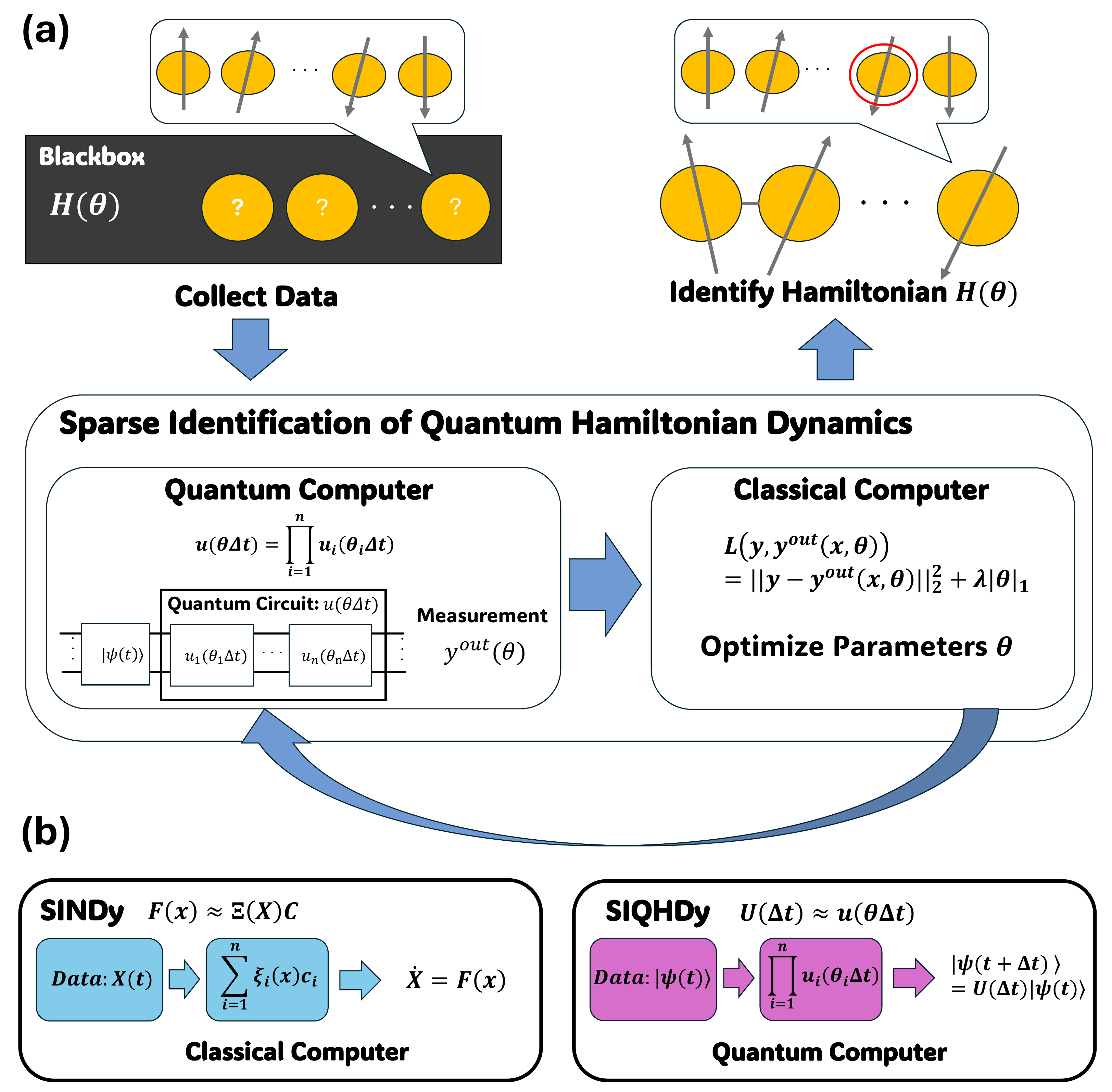}
			\caption{
                    (a) Schematic diagram of SIQHDy.  
                    (b) Correspondence between SINDy and  SIQHDy.
					}
			\label{fig1}
		\end{center}
\end{figure}
In this study, we propose sparse identification of quantum Hamiltonian dynamics (SIQHDy), a SINDy-inspired quantum circuit learning framework for estimating quantum Hamiltonian dynamics from time-series measurement data. 
A schematic diagram of the SIQHDy framework and its correspondence with SINDy is shown in Fig.~\ref{fig1}.
In SIQHDy, the Hamiltonian is expressed as a product of a predefined set of basis quantum circuits, and the corresponding circuit parameters are estimated through sparsity-promoting optimization via quantum circuit learning. This framework enables simultaneous identification of both the structure and parameters of the underlying quantum Hamiltonian dynamics.
We numerically demonstrate that SIQHDy successfully reconstructs the dynamics of single-, three-, and five-spin systems, including the transverse-field Ising model.
We further investigate the influence of measurement noise in the three-spin case and show that incorporating sparsity improves robustness to measurement noise and yields simple Hamiltonian models that accurately capture the underlying quantum dynamics. 
Moreover, we propose an extension of SIQHDy for scenarios with limited accessible observables and validate its performance in identifying two-spin systems and in reconstructing the network structures of five-spin systems.

It is worth noting that the compressive-sensing-based Hamiltonian estimation method~\cite{shabani2011estimation} may also be applied to the sparse identification of quantum Hamiltonian dynamics and can be viewed as a variant of compressive-sensing-based sparse identification methods for classical nonlinear dynamics~\cite{wang2011predicting}.
In contrast to this previous study, the proposed SIQHDy not only sparsely identifies the quantum Hamiltonian dynamics but also directly constructs a quantum circuit simulator that approximately reproduces the original quantum Hamiltonian dynamics.
%
%
\section{Overview of SINDy and QCL}
The proposed SIQHDy framework is a quantum circuit learning algorithm inspired by the SINDy method in classical nonlinear dynamics. 
In this section, to clarify its conceptual foundations, we provide a brief overview of sparse identification of nonlinear dynamics (SINDy) \cite{brunton2016discovering} and 
quantum circuit learning (QCL) \cite{mitarai2018quantum}.

\subsection{Sparse identification of nonlinear dynamics (SINDy)}

SINDy~\cite{brunton2016discovering} is a data-driven framework for estimating nonlinear dynamical systems from time-series data. In this framework, the system dynamics is approximated as a linear combination of a predefined set of basis functions, with their coefficients sparsely estimated from the time-series data. Owing to this sparsity, SINDy simultaneously identifies both the governing equations and the underlying dominant structures of the nonlinear dynamical system.

We consider a dynamical system,
\begin{align}
	\label{eq:X}
	\dot {\bm x}(t) = \bm{F}(\bm{x}(t)),
\end{align}
where $\bm x(t) \in{\mathbb R}^{n \times 1}$ represents the system state at time $t$, $\bm{F}(\bm{x}(t)) \in {\mathbb R}^{n \times 1}$ denotes a smooth vector field describing its dynamics, and $(\dot{})$ indicates the time derivative.

We sample the system state in Eq.~(\ref{eq:X}) at uniform intervals $\Delta t$, namely,
$\bm{x}_k = \bm{x}(k \Delta t)~(k = 0, 1, \dots, m)$. 
We construct a data matrix
\begin{align}
	\label{eq:dat1}
	\bm{X}=\left[\bm{x}_{0}, \bm{x}_{1}, \cdots, \bm{x}_{m-1} \right]^\mathsf{T},
\end{align}
where $^\mathsf{T}$ denotes matrix transpose.
We also construct a data matrix for the time derivative of the system state using an appropriate finite-difference method, e.g. $\dot{\bm{x}}_{k} = (\bm{x}_{k+1} - \bm{x}_{k})/\Delta t$,
yielding,
\begin{align}
	\label{eq:dat2}
	\dot{\bm{X}}=\left[ \dot{\bm{x}}_{0}, \dot{\bm{x}}_{1}, \cdots, \dot{\bm{x}}_{m-1} \right]^\mathsf{T}.
\end{align}

We also introduce a library of basis functions
$ \bm{\xi}(\bm{x})=[\xi_1(\bm{x}),\xi_2(\bm{x}),\cdots,\xi_l(\bm{x})]^\mathsf{T}  \in {\mathbb R}^{l \times 1}$ 
together with an associated coefficient vector 
$\bm{c}_j \in  {\mathbb R}^{l \times 1}~(j = 1, 2, \ldots, n)$. 
Introducing the coefficient matrix 
$\bm{C} = [\bm{c}_1, \bm{c}_2,\dots,\bm{c}_n]  \in {\mathbb R}^{l \times n}$, 
we approximate the vector field $\bm{F}$ as,
\begin{align}
	\bm{F}(\bm{x}) \approx   \bm{C}^\mathsf{T}  \bm{\xi}(\bm{x}).
\end{align}
Accordingly, we construct the basis function matrix 
$\bm{\Xi}(\bm{X}) = [\bm{\xi}(\bm{x}_0), \bm{\xi}(\bm{x}_1), \cdots, \bm{\xi}(\bm{x}_{m-1})]^\mathsf{T} \in {\mathbb R}^{m \times l}$
and consider the following approximation of the underlying dynamical system 
\begin{align}
	\dot{\bm{X}} \approx \bm{\Xi}(\bm{X}) \bm{C}.
\end{align}
The SINDy framework assumes that the vector field $\bm{F}$ can be accurately approximated by a small number of dominant basis functions. Under this assumption, the coefficient vectors $\bm{c}_j$
comprising the matrix 
$\bm{C}$ are estimated by solving the following sparse optimization problem
\begin{align}
	\label{eq:argmin_A}	
	\bm{c}_j  = \argmin_{\bar{\bm{c}}_j}  \left\| \dot{\bm{X}}_j - \bm{\Xi}(\bm{X}) \bar{\bm{c}}_j   \right\|^2_{2} +  \lambda \left\| \bar{\bm{c}}_j \right\|_{1},
\end{align}
where $\dot{\bm{X}}_j$ denotes the $j$-th column of $\dot{\bm{X}}$, $\left\|  \cdot \right\|_{1}$ and $\left\|  \cdot \right\|_{2}$ denote the $\ell_{1}$- and $\ell_{2}$- norms, respectively, and $\lambda$ is a regularization parameter. 

In practice, rather than explicitly including the  $\ell_1$-norm regularization term in Eq.~(\ref{eq:argmin_A}), the sequential thresholded least-squares algorithm~\cite{brunton2016discovering} 
is employed to promote sparsity due to its computational efficiency. 
This sparsity-promoting regression enables simultaneous estimation of the underlying dynamics and identification of the dominant basis functions.

\subsection{Quantum Circuit Learning}
\label{s2.1}
Quantum circuit learning (QCL)~\cite{mitarai2018quantum} is a hybrid quantum-classical machine learning algorithm developed for noisy intermediate-scale quantum (NISQ) devices. In QCL, classical data is encoded into quantum states and processed by parameterized quantum circuits, while circuit parameters are trained using classical optimization algorithms running on classical computers based on quantum measurement outcomes. These adjustable circuit parameters play a role analogous to trainable weights in classical neural networks. By appropriately tuning the parameters, QCL can be applied to a wide range of tasks, including function approximation, classification, and quantum simulation~\cite{mitarai2018quantum}.
The QCL algorithm is summarized as follows:

\begin{enumerate}
    \item Data encoding \mbox{}\\
    We encode the input data $\bm{x}$
    into a quantum state by applying a data-dependent unitary gate $V(\bm{x})$, producing the following input state,
    \begin{align}
        \ket{\psi_{in}(\bm{x})}=V(\bm{x})\ket{0}.
    \end{align}

    \item Parameterized quantum circuit \mbox{}\\
    We evolve the input state by applying a $\bm{\theta}$-parameterized quantum circuit $U(\bm{\theta})$, producing the output state,
    \begin{align}
        \ket{\psi_{out}(\bm{x}, \bm{\theta})}=U(\bm{\theta})\ket{\psi_{in}(\bm{x})}.
    \end{align}
    \item Quantum measurement \mbox{}\\
    We introduce a set of $n_m$ measurement operators $M = [M_1, M_2, \dots, M_{n_m}]$ and obtain their expectation values through quantum measurements on the output state,
    $$\bm{y}^{out}(\bm{x}, \bm{\theta}) =
        [y^{out}_1(\bm{x}, \bm{\theta}), y^{out}_2(\bm{x}, \bm{\theta}), \dots, y^{out}_{n_m}(\bm{x}, \bm{\theta})],$$ 
    where each output component is given by,
    \begin{align}
    \begin{split}
        y^{out}_{j}(\bm{x}, \bm{\theta}) = \bra{\psi_{out}(\bm{x}, \bm{\theta})}M_j\ket{\psi_{out}(\bm{x}, \bm{\theta})}~(j = 1, 2, \dots, n_m).
    \end{split}
    \end{align}
    \item Parameter optimization  \mbox{}\\
    We update parameters $\bm{\theta}$ via a classical optimization algorithm to minimize the cost function $L( \bm{y}, \bm{y}^{out}(\bm{x}, \bm{\theta}))$, where $\bm{y}$ denotes the target output values.
\end{enumerate}

By iterating steps 2-4 until the cost function converges to a sufficiently small value, we obtain the optimal parameters $\bm{\theta}_{opt}$, enabling the quantum circuit to learn the desired input-output relationship.

\section{SIQHDy algorithms}
\label{s:3}
In this study, we propose sparse identification of quantum Hamiltonian dynamics (SIQHDy), a SINDy-inspired quantum circuit learning framework for sparsely identifying quantum Hamiltonian dynamics.

We consider the time evolution of an \(N\)-qubit quantum state governed by the Schrödinger equation,
\begin{align}
\label{eq:qsys}
	i \dot{\ket{\psi(t)}}= H \ket{\psi(t)},
\end{align}
where \(\ket{\psi(t)} \in \mathbb{C}^{2^N}\) denotes the system state at time \(t\), $H \in {\mathbb C}^{2^N \times 2^N}$ 
is the system Hamiltonian, 
$i = \sqrt{-1}$, and the reduced Planck constant is set as $ \hbar = 1$.

In SIQHDy, we aim to reconstruct the unitary time evolution over a finite-time interval $\Delta t$ starting from time $t$
\begin{align}
    \label{eq:qsys_uni}
	\ket{\psi(t+\Delta t)}= U(\Delta t) \ket{\psi(t)} = 
    e^{-iH \Delta t} \ket{\psi(t)}.
\end{align}
To this end, we perform quantum measurements on a complete set of orthogonal operators. In the case of a single-qubit system, we use the four basis operators \(\{G_k\}_{k=1}^{4} = \{X, Y, Z, I\}\), where $X, Y$ and $Z$ are the Pauli matrices and \(I\) denotes the identity operator. For an \(N\)-qubit quantum system, 
this construction is extended to a complete set of \(4^N\) orthogonal basis operators 
$$
\{ F_k \}_{k = 1}^{4^N} = \{  G_{k_1} \otimes G_{k_2} \otimes \cdots \otimes G_{k_n}  \}_{k_1, k_2, \cdots, k_n = 1,2,3,4}.$$
The corresponding measurement outcomes at time $t$ are
given by the expectation values 
$$\bm{m}(t) = [m_1(t), m_2(t), \cdots, m_{4^N}(t)],$$
where each component is given by 
$$m_k(t) = \bra{\psi(t)} F_k \ket{\psi(t)}\quad(k = 1, 2, \ldots, 4^{N} ).$$ 
In SIQHDy, the pair of measurement values $[\bm{m}(t), \bm{m}(t+ \Delta t)]$ is used as the input-output pair for quantum circuit learning.

To learn the unitary dynamics of the system described by Eq.~(\ref{eq:qsys_uni}), we construct a parameterized quantum circuit composed of a product of basis quantum circuits that approximates the unitary time-evolution operator,
\begin{align}
	 u(\bm{\theta}\Delta t) 
	=	\prod^{n}_{i=1}u_i(\theta_i \Delta t)
    \approx U(\Delta t).
\end{align}
Here, $u_i(\theta_i \Delta t)$ denotes the $i$-th basis quantum circuit and 
 \(\bm{\theta} = [\theta_1, \theta_2, \dots, \theta_n] \in \mathbb{R}^{n}\) is a set of circuit parameters characterizing the circuit operations.  These parameters \(\bm{\theta}\) are optimized via quantum circuit learning.

Figure~\ref{fig2} shows a schematic diagram of the SIQHDy algorithm, and the procedure is summarized as follows:
\begin{figure} [!t]
	\begin{center}
	\includegraphics[width=0.75\hsize,clip]{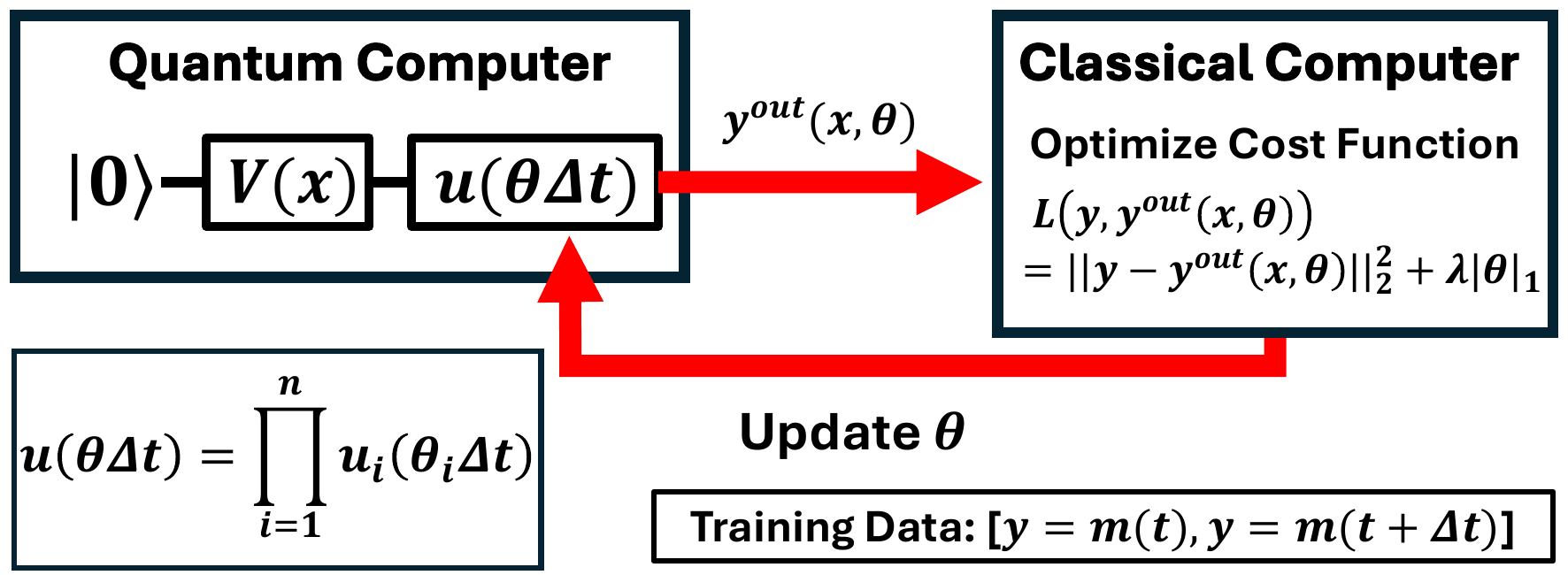}
			\caption{
                    A schematic diagram of the SIQHDy algorithm.
					}
			\label{fig2}
		\end{center}
\end{figure}
\begin{enumerate}
\item Data encoding \mbox{}\\
The initial input state is prepared as the quantum state \(\ket{\psi(t)}\) at time \(t\). Accordingly, the input data \(\bm{x} = \bm{m}(t)\) are encoded into the input state as 
	\begin{align}
		\ket{\psi_{in}(\bm{x})}= \ket{\psi(t)}, 
		\quad \ket{\psi(t)}\bra{\psi{(t)}}=  \sum_{k=1}^{4^N} \frac{1}{2^N}m_k(t) F_k.
        \label{eq:psi_in}
	\end{align}		

\item Parameterized quantum circuit \mbox{}\\
The parameterized quantum circuit \(u(\bm{\theta}\Delta t)\) is applied to the input state, yielding the output state,
\begin{align}
	\ket{\psi_{out}(\bm{x}, \bm{\theta})}=u(\bm{\theta} \Delta t)\ket{\psi_{in}(\bm{x})}.
\end{align}
\item Quantum measurement \mbox{}\\
The expectation values of the measurement operators \(M = \{ F_1, F_2, \cdots, F_{4^N} \}\) are obtained by performing quantum measurements on the output state, yielding 
$$\bm{y}^{out}(\bm{x}, \bm{\theta}) = 
[y^{out}_1(\bm{x}, \bm{\theta}), y^{out}_2(\bm{x}, \bm{\theta}), \cdots, y^{out}_{4^N}(\bm{x}, \bm{\theta})],$$
where each output component is given by 
\begin{align}
\begin{split}
	y^{out}_{j}(\bm{x}, \bm{\theta})=\bra{\psi_{out}(\bm{x}, \bm{\theta})}F_j\ket{\psi_{out}(\bm{x}, \bm{\theta})} \quad (j = 1, 2, \cdots, 4^N).
\end{split}
\end{align}
 \item Parameter optimization $\bm{\theta}$ \mbox{}\\
The parameters are sparsely selected so that the output values 
$\bm{y}^{out}(\bm{x}, \bm{\theta})$ match the target data $\bm{y} = \bm{m}(t+\Delta t)$.
To this end, we introduce the following cost function,
\begin{align}
	\label{eq:cst}
    \hspace{-0.5em}
	L(\bm{y}, \bm{y}^{out}(\bm{x}, \bm{\theta}))
	=  \left\| \bm{y} - \bm{y}^{out}(\bm{x}, \bm{\theta})  \right\|_{2}^{2} +  \lambda \left\| \bm{\theta} \right\|_{1},
\end{align}
where the $\ell_1$-norm regularization promotes sparsity in the parameters $\bm{\theta}$. The parameters are updated by minimizing this cost function using standard classical optimization algorithms.
In practical numerical calculations, sparsity is promoted without explicitly introducing the $\ell_1$-norm regularization term. Instead, after each parameter update obtained by minimizing the cost function without $\ell_1$-norm regularization, parameters whose absolute values are smaller than a prescribed threshold 
$\lambda$ are set to zero (see Appendix A for details).

\end{enumerate}
By iterating steps 2-4 until the cost function converges to a sufficiently small value, we obtain the optimal parameters $\bm{\theta}_{opt}$ such that $u(\bm{\theta}_{opt}\Delta t)  \approx U(\Delta t)$.

As in the numerical examples presented below, measurement data are typically collected at uniform sampling interval $\Delta t$, yielding a sequence of input-output pairs $[\bm{m}(k \Delta t), \bm{m}((k+1)\Delta t)]$ for $k = 0, 1, 2, \ldots$. Thus, for parameter optimization, the cost function can also be defined as the sum of Eq.~(\ref{eq:cst}) evaluated over multiple such input-output pairs.



\section{Numerical results}
As numerical demonstrations, we apply SIQHDy to single-, three-, and five-spin Hamiltonian systems and investigate the effect of measurement noise in the three-spin case.

We consider the time evolution of an \(N\)-qubit quantum state 
governed by Eq.~(\ref{eq:qsys}), where the underlying Hamiltonian is unknown. 
We consider quantum Hamiltonian systems with one- and two-body interactions, for which the interaction Hamiltonian is assumed to be sparse. This assumption is physically reasonable because realistic quantum systems are typically governed by a limited number of local one- and two-body interactions, while higher-order effective interactions are suppressed and become weaker with increasing distance or energy separation~\cite{nielsen2000quantum}. Consequently, the effective Hamiltonian can often be represented using only a small subset of all possible one- and two-body interaction terms.

Accordingly, for the parameterized quantum circuits, we construct basis quantum circuits as products of rotation gates acting on single-spin and two-spin subsystems selected from the $N$-qubit system. This construction relies on the assumption that the time evolution over a short sampling interval can be approximately represented as a product of such local rotation gates.
When the time interval $\Delta t$ in Eq.~(\ref{eq:qsys_uni}) is sufficiently small,
this assumption is justified by the Suzuki--Trotter decomposition
up to errors of the order $\mathcal{O}(\Delta t^2)$, and different gate orderings lead to equivalent dynamics within this approximation~\cite{nielsen2000quantum}.
\begin{figure} [!t]
	\begin{center}
			\includegraphics[width=1.0\hsize,clip]{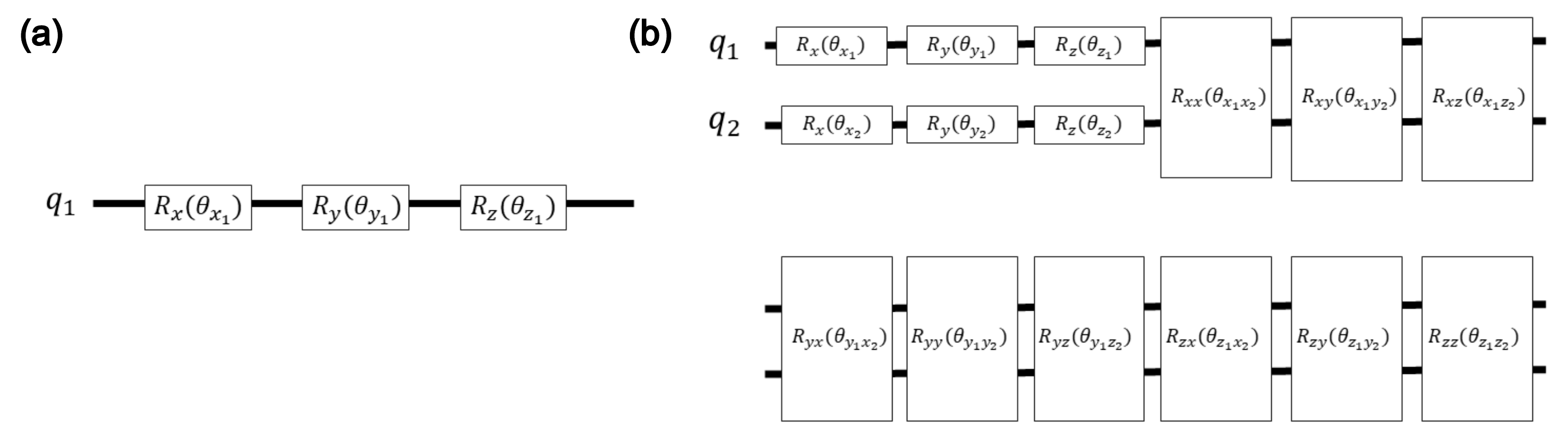}
    \caption{
    Basis quantum circuits for the single-qubit case (a) and the two-qubit case (b).
    (a) Three single-qubit rotation gates
    about $X$, $Y$, and $Z$ axes.
    (b) Fifteen rotation gates, consisting of $3 \times 2$ rotation gates acting on single qubits and $3 \times 3$ rotation gates acting on two qubits.
    Here, $q_1$ and $q_2$ denote the first and second qubits, respectively.
    }
			\label{fig3}
		\end{center}
\end{figure}
Figure~\ref{fig3} illustrates examples of basis quantum circuits for the single- and two-spin systems. In the single-spin case, three rotation gates are considered as candidates. In the two-spin case, $15$ candidates are considered, consisting of $3 \times 2$ rotation gates acting on single spins and $3 \times 3$ rotation gates acting on two spins. For systems with $N \geq 3$ spins, basis quantum circuits are constructed by choosing either one or two spins among the $N$ spins. Accordingly, the total number of candidate basis quantum circuits is given by $3 \times N + 3 \times 3 \times {}_N C_2$.

We collect the expectation values of the complete set of \(4^N\) orthogonal operators $\bm{m}(t)$ at discrete time steps
 \( t = k \Delta t~( k = 0, \ldots, 1000) \), with a sampling interval of \( \Delta t = 0.01 \), starting from a randomly chosen initial state.   
 We use the data at \( t = k \Delta t~( k = 0, \ldots, 100) \) for learning, and the remaining data are used to validate the learned quantum circuit.
 The data are collected by performing numerical simulations using the QuTiP numerical toolbox \cite{johansson2012qutip, johansson2013qutip}.
 
We implement quantum circuit learning using the Qiskit simulator in Python. At each optimization step, the circuit parameters are updated by minimizing the sum of the cost function in Eq.~(\ref{eq:cst}) over all input-output pairs  
$[\bm{m}(k \Delta t), \bm{m}((k+1)\Delta t)]$
sampled at discrete times with
$k = 0, 1, \ldots, 99$. 
Parameter optimization is performed using the COBYLA method provided by SciPy.

\subsection{A single-spin system}
As a simple illustrative example of the SIQHDy framework, we first consider a single-spin system characterized by the Hamiltonian
\begin{align}
H=\frac{3}{2}Y.
\label{eq:hm_single}
\end{align}
As basis quantum circuits, we prepare 
all three single-qubit rotation gates 
$R_x(\theta_{x_1}) = \exp{(-i\frac{\theta_{x_1}}{2}X)}, R_y(\theta_{y_1}) = \exp{(-i\frac{\theta_{y_1}}{2}Y)}, R_z(\theta_{z_1}) = \exp{(-i\frac{\theta_{z_1}}{2}Z)}$.
The parameters $\theta_{x_1} , \theta_{y_1}$ and $\theta_{z_1}$, corresponding to the rotation gates $X, Y$ and $Z$, respectively, are optimized so that the resulting quantum circuit reproduces the dynamics described by Eq.~(\ref{eq:qsys}) with the Hamiltonian in Eq.~(\ref{eq:hm_single}). As an illustrative example, 
when the Hamiltonian is given by $H = a X$, the corresponding unitary evolution over a time interval $\Delta t$ is given by $U(\Delta t) = \exp{(-i a \Delta tX)}$. Therefore, the estimated parameter is expected to satisfy $\theta_{x1} = a \cdot (2\Delta t)$.

\begin{figure} [!t]
	\begin{center}
			\includegraphics[width=1\hsize,clip]{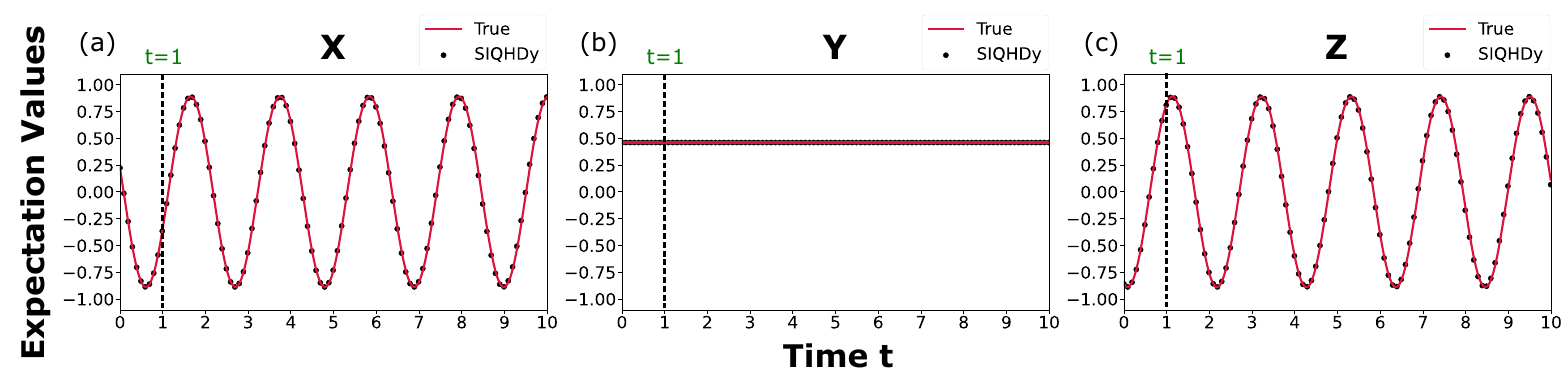}
			\caption{
            Results of the single-spin case.
            (a-c) Comparison between the true dynamics (red solid lines) of the original quantum system and the dynamics predicted by SIQHDy (black dots) for the expectation values of observables $X$ (a), $Y$ (b), and $Z$ (c).
            The dashed vertical line indicates $t=1$, up to which 
            the data are used for learning. 
            For $t \leq 1$ single-step predictions are plotted, whereas for $t > 1$ the learned quantum circuit is recursively applied starting from $t = 1$, and the resulting predicted dynamics are plotted.
            The SIQHDy results are plotted every $10$ sampling points.
            }
			\label{fig4}
		\end{center}
\end{figure}

In this case, the threshold parameter for the sparsity-promoting procedure in the SIQHDy algorithm is set to $\lambda = 0.05 \cdot (2\Delta t)$. 
As a result, the optimized circuit parameters are obtained as
$$
\bm{\theta}_{\mathrm{opt}}=[\theta_{x_1} , \theta_{y_1} , \theta_{z_1} ] = [0,1.503,0]\cdot (2\Delta t),$$ 
yielding a cost function value close to zero, namely $7.058 \times 10^{-7}$, confirming the accuracy of the estimation.

Figure~\ref{fig4} compares the true dynamics of the original quantum system in Eq.~(\ref{eq:qsys}) with those predicted by the learned quantum circuit. For $t \leq 1$, we evaluate one-step predictions using the measurement data at each sampling time.
For $t > 1$, starting from the measurement data at $t = 1$, we recursively apply the learned quantum circuit to predict the dynamics at subsequent sampling times. 
These results demonstrate that the learned quantum circuit accurately reproduces the original dynamics of the expectation values of observables $X$, $Y$, and $Z$ even for $t > 1$, after the learning interval.

\subsection{Three-spin system}
\label{sec:three_qubit}

Next, we consider a three-spin system characterized by the Hamiltonian
$$
H = \frac{3}{2}(XXI+ZZI)+IXX+IZZ.
$$

As basis quantum circuits, we prepare all rotation gates acting on single-qubit and two-qubit subsystems. The total number of basis quantum circuits is $3 \times 3 + 3 \times 3 \times {}_3 C_2 =  36$.
For example, the two-qubit rotation gate $XYI$
acting on the first and second qubits
is represented by $R_{xy}(\theta_{x_1y_2}) = \exp{(-i\frac{\theta_{x_1y_2}}{2}X \otimes Y)} \otimes I $.

In this case, the threshold parameter for the sparsity-promoting procedure in the SIQHDy algorithm is set to $\lambda = 0.25 \cdot (2\Delta t)$.  As a result,  the optimized circuit parameters with nonzero values are obtained as
$$
\bm{\theta}_{\mathrm{opt}}=
  [\theta_{x_{1}x_{2}}, \theta_{z_{1}z_{2}} ,\theta_{x_{2}x_{3}}, \theta_{z_{2}z_{3}}] = [1.502, 1.502, 1.006, 0.997] \cdot (2 \Delta t),
$$
corresponding to the two-qubit rotation gates, $XXI$, $ZZI$, $IXX$, and $IZZ$, respectively.
These estimated parameters yield a cost function value close to zero, namely 
$8.445 \times 10^{-7}$,  
confirming the accuracy of the estimation.
\begin{figure} [!t]
	\begin{center}
			\includegraphics[width=0.7\hsize,clip]{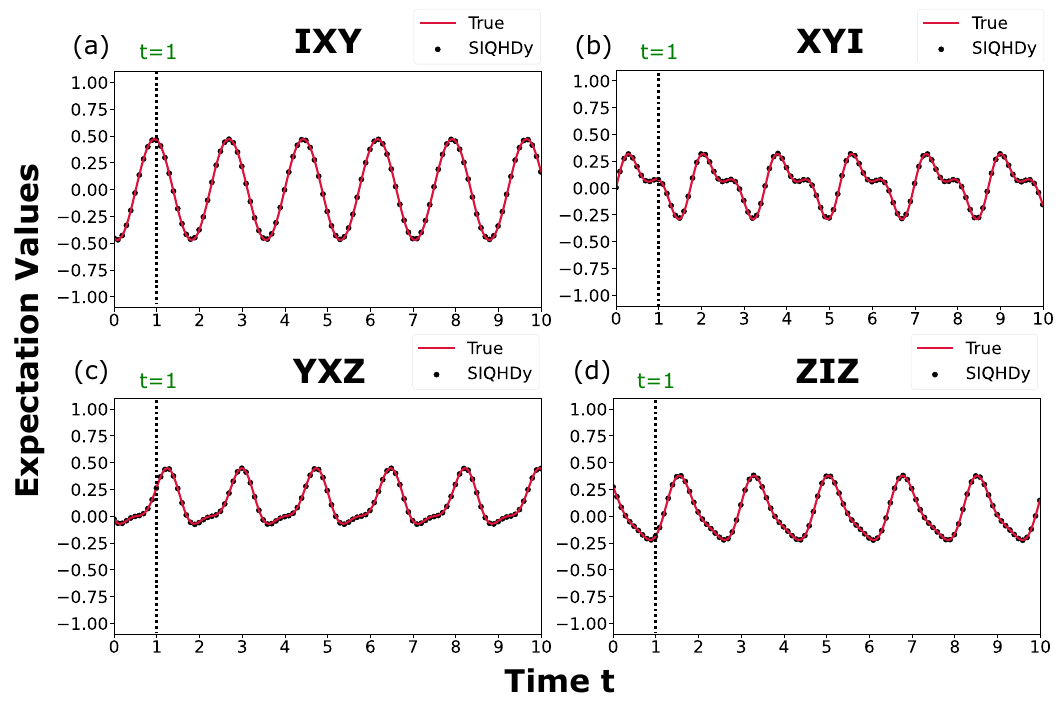}
			\caption{
            Results of the three-spin case. (a-d) Comparison between the true dynamics (red solid lines) of the original quantum system and the dynamics predicted by SIQHDy (black dots) for the expectation values of observables $IXY$ (a), $XYI$ (b), $YXZ$ (c), and $ZIZ$ (d).
            The dashed vertical line indicates $t=1$, up to which 
            the data are used for learning. 
            For $t \leq 1$ single-step predictions are plotted, whereas for $t > 1$ the learned quantum circuit is recursively applied starting from $t = 1$, and the resulting predicted dynamics are plotted.
            The SIQHDy results are plotted every $10$ sampling points.     
            }
			\label{fig5}
		\end{center}
\end{figure}

Figure~\ref{fig5} compares the true dynamics of the original quantum system in Eq.~(\ref{eq:qsys}) with those predicted by the learned quantum circuit, in a manner similar to the previous case. It can be seen that the learned quantum circuit accurately predicts the original dynamics of the expectation values of observables $IXY$, $XYI$, $YXZ$, and $ZIZ$.

\subsection{Five-spin system}
Next, we consider a five-spin transverse-field Ising model characterized by the Hamiltonian
\begin{align*}
\begin{split}
H=\sum_{i=1}^{5}X_{i}+\sum_{j=1}^{4}J_{j}Z_{j}Z_{j+1},\quad 
(J_1, J_2, J_3, J_4) = (2.5,2.0,1.5,1.0).\\
\end{split}
\end{align*}
As basis quantum circuits, we prepare rotation gates acting on single-qubit and two-qubit subsystems.
The total number of basis quantum circuits is 
$3 \times 5 + 3 \times 3 \times {}_5 C_2 =  105$. 
For example, the two-qubit $XY$ rotation gate acting on the second and third qubits is represented by 
$R_{x_2y_3}(\theta_{x_2y_3}) = I \otimes \exp{(-i\frac{\theta_{x_2y_3}}{2}X \otimes  Y)}\otimes I \otimes I$.

In this case, the threshold parameter for the sparsity-promoting procedure in the SIQHDy algorithm is set to $\lambda = 0.25\cdot (2\Delta t)$.
As a result of the SIQHDy method, the optimized circuit parameters with nonzero values are obtained as 
\begin{align*}
\bm{\theta}_{\mathrm{opt}}&=
  [\theta_{x_{1}}, 
  \theta_{x_{2}},
  \theta_{x_{3}},
  \theta_{x_{4}},
  \theta_{x_{5}},
  \theta_{z_{1}z_{2}}, 
  \theta_{z_{2}z_{3}}, 
  \theta_{z_{3}z_{4}},
  \theta_{z_{4}z_{5}}
  ] \\
  &=[0.999,\,1.002 ,\,1.002 ,\, 0.998,\,1.000   
  ,\,2.502,\,2.002,\,1.497,\,  1.003],
\end{align*}
corresponding to rotation gates
$XIIII$, $IXIII$, $IIXII$, $IIIXI$, $IIIIX$, $ZZIII$, $IZZII$, $IIZZI$, and 
$IIIZZ$, respectively.
These estimated parameters  
result in a cost function value close to zero, namely
$1.293 \times 10^{-6}$, 
confirming the accuracy of the estimation. 
\begin{figure} [!t]
	\begin{center}
			\includegraphics[width=0.7\hsize,clip]{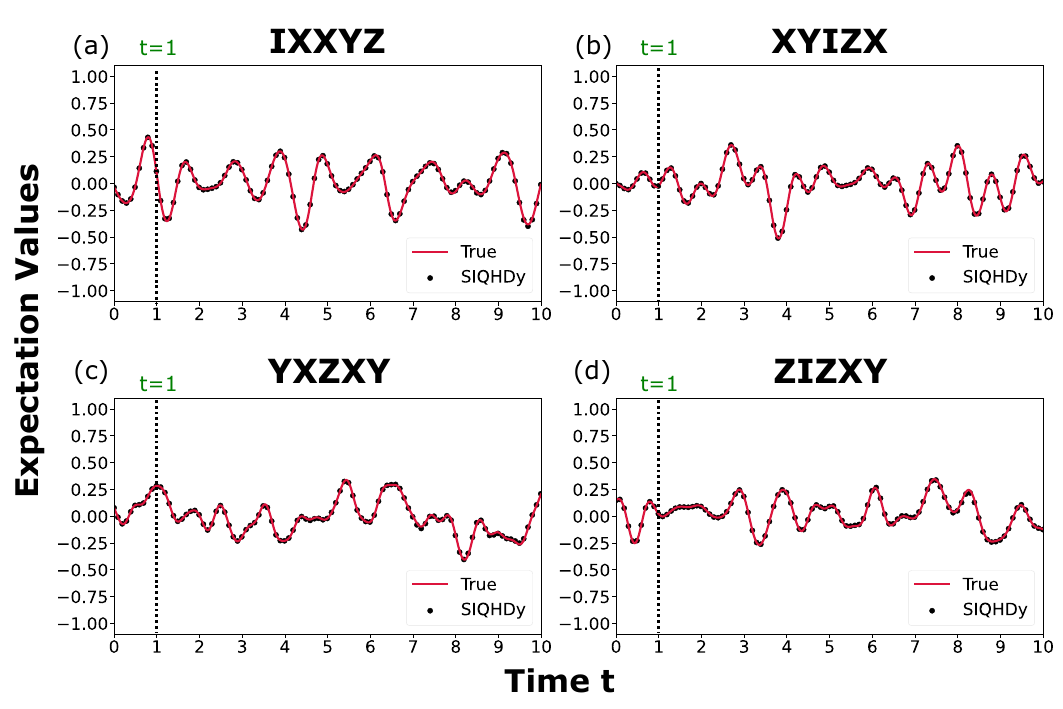}
			\caption{
            Results of the five-spin case. (a-d) Comparison between the true dynamics (red solid lines) of the original quantum system and the dynamics predicted by SIQHDy (black dots) for the expectation values of observables $IXXYZ$ (a), $XYIZX$ (b), $YXZXY$ (c), $ZIZXY$ (d).
            The dashed vertical line indicates $t=1$, up to which 
            the data are used for learning. 
            For $t \leq 1$ single-step predictions are plotted, whereas for $t > 1$ the learned quantum circuit is recursively applied starting from $t = 1$, and the resulting predicted dynamics are plotted.
            The SIQHDy results are plotted every $10$ sampling points.         
            }
			\label{fig6}
		\end{center}
\end{figure}

Figure~\ref{fig6} compares the true dynamics of the original quantum system in Eq.~(\ref{eq:qsys}) with those predicted by the learned quantum circuit, in a manner similar to the previous case. 
The results demonstrate that 
the expectation values of observables $IXXYZ$, $XYIZX$, $YXZXY$, and $ZIZXY$ are accurately reproduced by the learned quantum circuit.

\subsection{Three-spin system with measurement noise}

In the previous examples, we considered ideal noiseless measurement data. Although the results are not shown, we numerically confirmed that accurate parameter estimation can be achieved in these cases even without sparse optimization, while very small residual coefficients may remain in the learned model.
By contrast, the advantages of sparse optimization become evident in the presence of measurement noise, as it enhances robustness to noise.

We consider the three-spin system introduced in Sec. IV B and investigate the effects of measurement noise. 
We assume that the quantum measurements are performed in a noisy environment such that the expectation values of the measurement operators are subject to additive Gaussian noise.
$$
m_j(k\Delta t) = \bra{\psi(k\Delta t)} F_j \ket{\psi(k\Delta t)} + \xi_j(k\Delta t), 
$$
where $\xi_j(k\Delta t) \sim \mathcal{N}(0, \sigma^2)$~$(j = 1, 2, \ldots, 4^3, k = 0, 1, \ldots, 1000)$ with $\sigma$ denoting the standard deviation of the measurement noise. 
We apply the SIQHDy method using these noisy measurement data. In the numerical simulations, we set $\sigma = 0.05$.

In this case, the threshold parameter for the sparsity-promoting procedure in the SIQHDy algorithm is set to $\lambda = 0.3 \cdot (2\Delta t)$. 
As a result of the SIQHDy method, the optimized circuit parameters with nonzero values are obtained as 
\begin{align*}
\bm{\theta}_{\mathrm{opt}}=
  [\theta_{x_{1}x_{2}}, \theta_{z_{1}z_{2}} ,\theta_{x_{2}x_{3}}, \theta_{z_{2}z_{3}} ]  
  = [1.522, 1.503,0.982,1.005 ] \cdot (2\Delta t),
\end{align*}
corresponding to rotation gates, $XXI, ZZI, IXX$, and $IZZ$, respectively.
The estimated parameters yield a cost function value close to zero, namely, 
$ 2.784\times 10^{-6}$, confirming the accuracy of the estimation. 

For comparison, we also perform SIQHDy without sparse optimization, that is, by minimizing the cost function 
in Eq.~(\ref{eq:cst}) without the $\ell_1$-norm regularization term at each optimization step. 
Without sparse optimization, the estimated parameters with magnitudes greater than the threshold $\lambda$ are obtained as
\begin{align*}
\bm{\theta}_{\mathrm{opt}}=
  [\theta_{x_{1}x_{2}}, \theta_{z_{1}z_{2}} ,\theta_{x_{2}x_{3}}, \theta_{z_{2}z_{3}} ]  
  = [1.598, 1.301,0.938, 1.160]\cdot (2\Delta t),
\end{align*}
corresponding to rotation gates $XXI, ZZI, IXX$, and $IZZ$, respectively.
The estimated parameters yield a cost function value close to zero, namely 
$9.076 \times 10^{-3}$,
confirming the accuracy of the estimation for the learning data.
However, without sparse optimization, small nonzero coefficients with magnitudes below the threshold $\lambda$
remain for other basis quantum circuits due to the measurement noise.

\begin{figure} [!t]
	\begin{center}
			\includegraphics[width=0.55\hsize,clip]{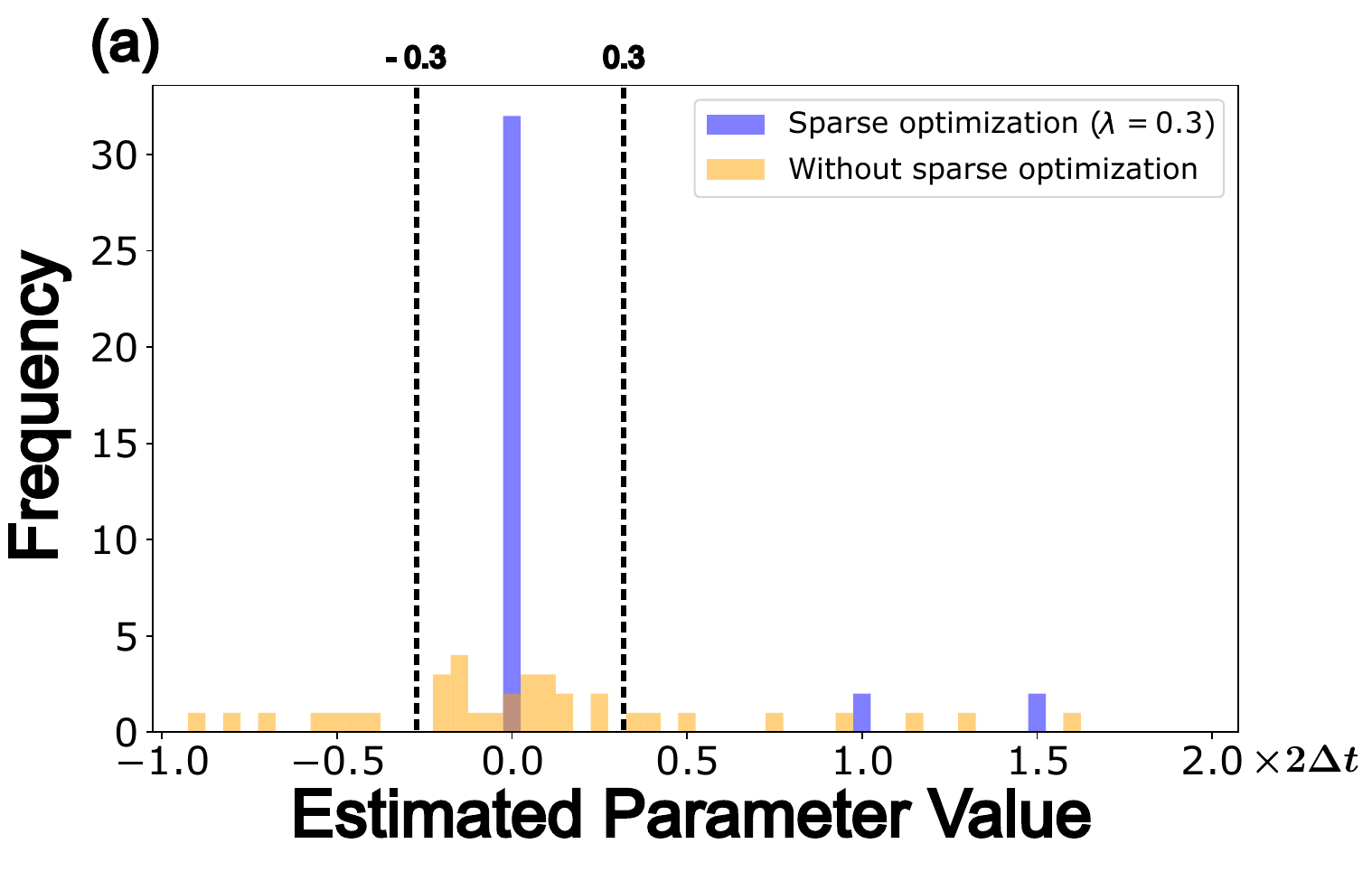}
			\caption{
            The blue and orange histograms correspond to the results with sparse optimization ($\lambda = 0.3$) and without sparse optimization, respectively. The dashed vertical lines indicate the threshold values $\pm 0.3$ applied to the circuit parameters in the sparsity-promoting procedure. The bin width of the estimated parameter values is $0.05$.
            }
			\label{fig7}
		\end{center}
\end{figure}

Figure~\ref{fig7} shows the distributions of the parameters estimated by SIQHDy with and without sparse optimization. Without sparse optimization, overfitting to noisy measurement data leads to small nonzero coefficients for other candidate gates. By contrast, sparse optimization eliminates these spurious coefficients, resulting in a simple and interpretable model.
To quantify this effect of measurement noise, we evaluate the parameter estimation error defined as
\begin{align*}
\mathrm{E}_{\mathrm{param}}
= \left\| \hat{\boldsymbol{\theta}} - \boldsymbol{\theta} \right\|_2 ,
\end{align*}

where $\boldsymbol{\theta}$ and  $\hat{\boldsymbol{\theta}}$
represent the true and estimated Hamiltonian parameters, respectively.
With sparse optimization, the estimation error is reduced to 
$\mathrm{E}_{\mathrm{param}} = 2.942 \times 10^{-2} \cdot (2 \Delta t)$, 
whereas without sparse optimization it remains larger at $\mathrm{E}_{\mathrm{param}} = 2.129 \cdot (2 \Delta t)$. 
%
This increase in error is mainly attributable to spurious small nonzero coefficients associated with other candidate gates, which arise due to the measurement noise. These results demonstrate that sparse optimization enhances robustness to measurement noise by suppressing such spurious coefficients and eliminating unnecessary basis quantum circuits.
\begin{figure} [!t]
	\begin{center}
			\includegraphics[width=1\hsize,clip]{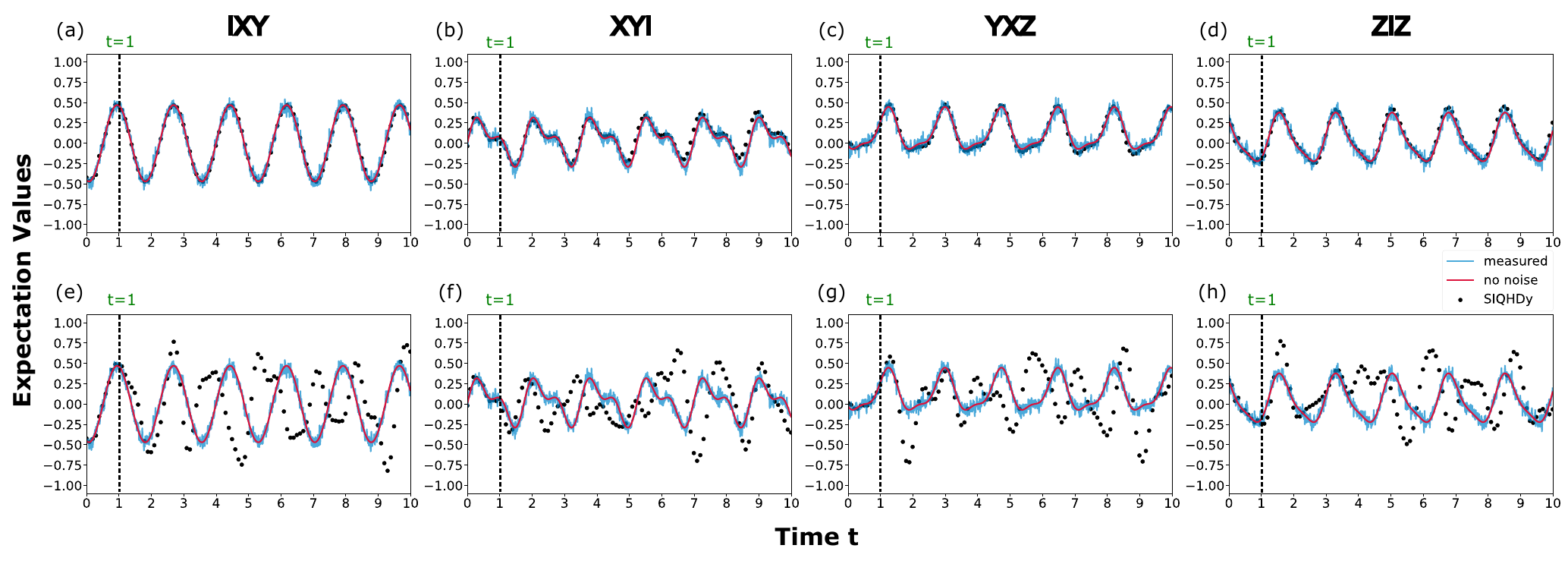}
			\caption{
            Results of the three-spin case with measurement noise.
            (a-h)
            Comparison between the true dynamics of the original quantum system without noise (red solid lines), the measured data with noise (blue solid lines), and the dynamics predicted by SIQHDy (black dots) for the expectation values of observables $IXY$ (a,e), $XYI$ (b,f), $YXZ$ (c,g), and $ZIZ$ (d,h). Panels (a-d) show the results with sparse optimization, whereas panels (e-h) show those without sparse optimization.
            The dashed vertical line indicates $t=1$, up to which 
            the data are used for learning. 
            For $t \leq 1$ single-step predictions are plotted, whereas for $t > 1$ the learned quantum circuit is recursively applied starting from $t = 1$, and the resulting predicted dynamics are plotted.
            The SIQHDy results are plotted every $10$ sampling points.  
            }
			\label{fig8}
		\end{center}
\end{figure}
Figure~\ref{fig8} compares the true dynamics of the original quantum system in Eq.~(\ref{eq:qsys}) with those produced by the learned quantum circuit, both with (a-d) and without (e-h) sparse optimization. The measured data and the true dynamics without noise are plotted together with the dynamics produced by the learned quantum circuit, which are obtained in a manner similar to the previous case.
In this case, we start from the noisy measurement data at $t = 1$ and recursively apply the learned quantum circuit to predict the dynamics at subsequent sampling steps.
As shown in Fig.~\ref{fig8}(a-d), the learned quantum circuit with sparse optimization accurately reproduces the true trajectories of the expectation values of observables $IXY$, $XYI$, $YXZ$, and $ZIZ$. By contrast, 
as shown in Fig.~\ref{fig8}(e-h), without sparse optimization, the learned quantum circuit fails to track the true trajectories due to the measurement noise.

\section{Extension of SIQHDy under Limited Measurement Access}

The SIQHDy framework has been formulated under the assumption that expectation values of all observables are accessible. In realistic quantum experiments, however, such complete measurement access is rarely available, as accessible measurements are often limited to a subset of qubits or observables due to hardware constraints and experimental overhead.

In this section, we develop an extension of the SIQHDy framework for scenarios with limited measurement access and validate its performance in identifying two-spin systems as well as reconstructing the network structures of five-spin systems.

\subsection{Algorithm}

In the data-encoding step of the SIQHDy algorithm, preparing the input quantum state \(\ket{\psi(t)}\) at time \(t\) in Eq.~(\ref{eq:psi_in}) assumes full access to measurement data for all observables. Under this assumption,  the unitary time evolution 
starting from time $t$ is approximated over a single sampling time step $\Delta t$. 
When measurement access is limited, however, preparing the system state at time $t$ becomes infeasible due to the lack of measurement data over a complete set of orthogonal basis operators.
Instead, starting from a known initial state prepared at $t = 0$,
we approximate the original unitary time evolution at later sampling times through repeated applications of a parameterized quantum circuit \cite{gupta2023hamiltonian}.

We assume that the initial state of the system is known, while only a limited set of measurement outcomes 
for the observables 
\(M' = \{F'_1, F'_2, \ldots, F'_{N'}\}\)
is accessible.
The corresponding measurement outcomes at time $t$ are
given by the expectation values 
$$\bm{m'}(t) = [m'_1(t), m'_2(t), \cdots, m'_{N'}(t)],\quad m'_j(t) = \bra{\psi(t)} F'_j \ket{\psi(t)}~(j = 1, 2, \ldots, N').$$ 
In the extended SIQHDy framework under limited measurement access, we use the known initial state $\ket{\psi(0)}$ at time $t = 0$ and the accessible measurement data $\bm{m}'(k \Delta t)$ at time $t = k \Delta t$ as an input-output pair $[\ket{\psi(0)}, \bm{m}'(k\Delta t)]$
for $k = 1, 2, \ldots$.
\begin{figure} [!t]
	\begin{center}
			\includegraphics[width=1.0\hsize,clip]{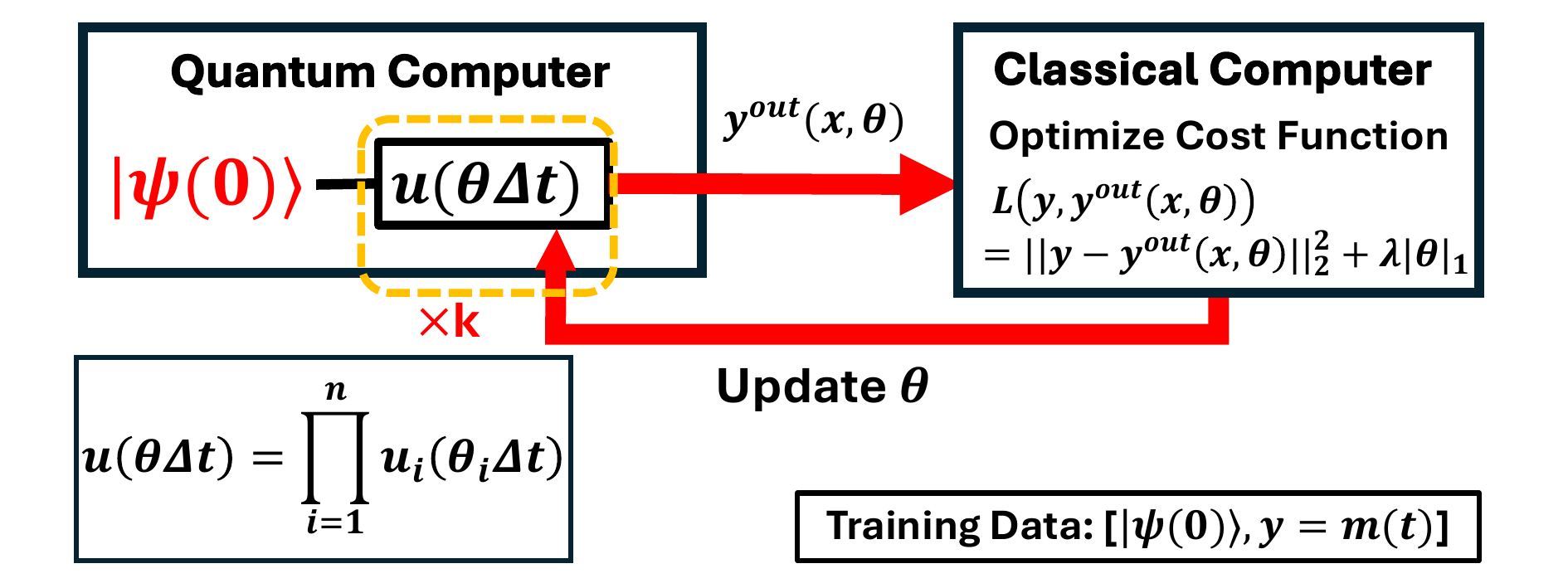}
			\caption{
                    A schematic diagram of the extended SIQHDy framework 
                    under limited measurement access. 
					}
			\label{fig9}
		\end{center}
\end{figure}

A schematic diagram of the extended SIQHDy framework under limited measurement access is shown in Fig.~\ref{fig9}, and the corresponding algorithm is summarized as follows.

\begin{enumerate}
\item Initial state encoding \mbox{}\\
The initial input state is prepared as the quantum state at time $t = 0$, which is assumed to be known and is used as the input data
\begin{align}
    \ket{\psi_{in}(\bm{x})}= \ket{\psi(0)}. 
\end{align}		

\item Parameterized quantum circuit \mbox{}\\
The parameterized quantum circuit 
\(u(\bm{\theta}\Delta t)\), which approximates the unitary evolution over a finite time interval
$\Delta t$, is repeatedly applied to the input state $k$ times, yielding the output quantum state at time $t = k \Delta t$
\begin{align}
	\ket{\psi_{out}(\bm{x}, \bm{\theta})}=u(\bm{\theta} (\Delta t))^k\ket{\psi_{in}(\bm{x})}.
\end{align}
\item Quantum measurement \mbox{}\\
The expectation values of the measurement operators
 \(M' = \{ F'_1, F'_2, \cdots, F'_{N'} \}\) are obtained by performing quantum measurements on the output state, yielding 
\(\bm{y}^{out}(\bm{x}, \bm{\theta}) = 
[y^{out}_1(\bm{x}, \bm{\theta}), y^{out}_2(\bm{x}, \bm{\theta}), \cdots, y^{out}_{N'}(\bm{x}, \bm{\theta})]\),
where each output component is given by 
\begin{align}
\begin{split}
	\bm{y}^{out}_{j}(\bm{x}, \bm{\theta})=\bra{\psi_{out}(\bm{x}, \bm{\theta})}F'_j\ket{\psi_{out}(\bm{x}, \bm{\theta})}~(j = 1, 2, \cdots, N').
\end{split}
\end{align}
 \item Parameter optimization $\bm{\theta}$ \mbox{}\\
 The parameters are sparsely selected so that the output values 
$\bm{y}^{out}(\bm{x}, \bm{\theta})$ match the target data $\bm{y} = \bm{m'}(k \Delta t)$.
To this end, we introduce the following cost function
\begin{align}
	\label{eq:cst2}
    \hspace{-0.5em}
	L(\bm{y}, \bm{y}^{out}(\bm{x}, \bm{\theta}))
	=  \left\| \bm{y} - \bm{y}^{out}(\bm{x}, \bm{\theta})  \right\|_{2}^{2} +  \lambda \left\| \bm{\theta} \right\|_{1},
\end{align}
where the $\ell_1$-norm regularization promotes sparsity in the parameters $\bm{\theta}$. The parameters are updated by minimizing this cost function using standard classical optimization algorithms. In the practical numerical implementation, sparsity is promoted using the same procedure as in the original SIQHDy formulation, as described in Appendix~A.
\end{enumerate}

By iterating steps 2-4 until the cost function converges to a sufficiently small value, we obtain the optimal parameters $\bm{\theta}_{\mathrm{opt}}$, which allow the quantum circuit to learn the desired input-output relationship.

As in the following examples, when measurement data are collected at uniform sampling intervals $\Delta t$
from multiple trajectories starting from different initial states, the cost function can be defined as the sum of Eq.~(\ref{eq:cst2}) evaluated over all output data $\bm{m'}(k \Delta t)$ at multiple sampling points $k = 1, 2, \ldots$ and over all trajectories.

\subsection{Numerical results}

We numerically demonstrate the validity of the extended SIQHDy framework under limited measurement access by identifying two-spin systems and reconstructing network structures of five-spin systems.

We collect the expectation values of the accessible measurement operators at discrete time steps \( t = k \Delta t~( k = 0, \ldots, 100) \), with a sampling interval of \( \Delta t = 0.01 \), starting from multiple randomly chosen initial states, each assumed to be known. 
We use the data at \( t = k \Delta t~( k = 0, \ldots, 10) \) from all trajectories for learning, while the remaining data from a single trajectory are used to validate the learned quantum circuit.

At each optimization step, the circuit parameters are updated by minimizing the sum of the cost function in Eq.~(\ref{eq:cst2}) over all data sampled at  $t = k \Delta t~(k = 0, 1, \ldots, 10)$
and over all trajectories. We collect trajectories from $3$ and $10$ different initial states for the two-spin and five-spin cases, respectively.

\subsubsection{Two-spin system}
\label{sec:three_qubit}

As a first example, we consider a two-spin system 
characterized by the Hamiltonian
$$
H = XX + ZZ.
$$
We assume that only the observables for the first spin are accessible
$$M' = \{ XI, YI, ZI \}$$
and measure their expectation values at the sampling times.

As basis quantum circuits, we prepare all rotation gates acting on single-qubit and two-qubit subsystems, shown in Fig.~\ref{fig3}(b).

In this case, the threshold parameter for the sparsity-promoting procedure in the extended SIQHDy algorithm is set to $\lambda = 0.25 \cdot (2\Delta t)$. As a result, the estimated parameters with nonzero values are obtained as
\begin{align*}
\bm{\theta}_{\mathrm{opt}}=
  [\theta_{x_{1}x_{2}}, \theta_{z_{1}z_{2}}] = [1.000, 1.000] \cdot (2\Delta t),
\end{align*}
corresponding to $XX$ and $ZZ$ rotation gates, respectively.
These estimated parameters yield a cost function value close to zero, namely 
$2.048 \times 10^{-12}$, 
confirming the accuracy of the estimation. 

\begin{figure} [!t]
	\begin{center}
			\includegraphics[width=1\hsize,clip]{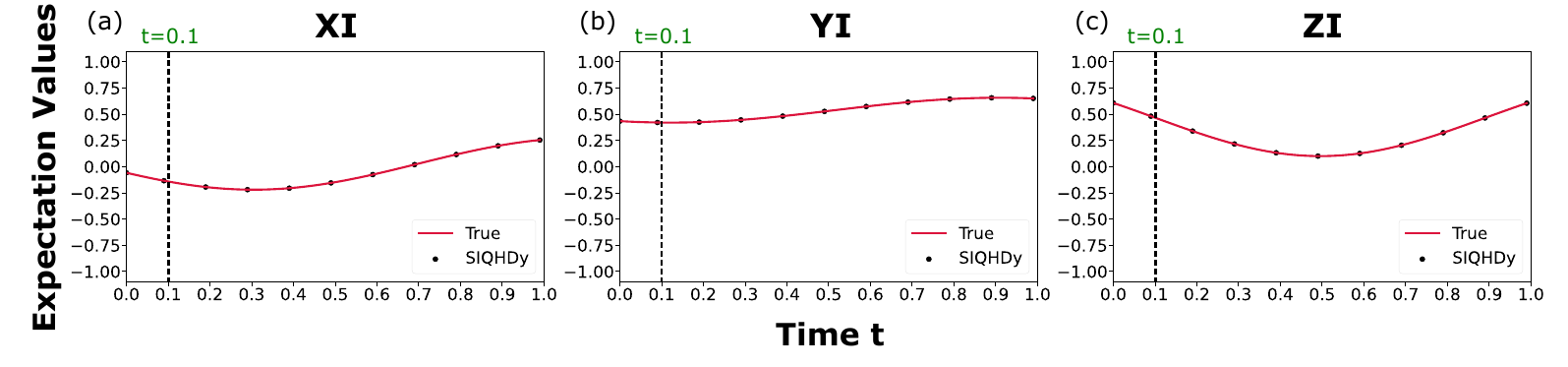}
			\caption{
            Results of the two-spin case.
            (a-c) Comparison between the true dynamics (red solid lines) of the original quantum system and the dynamics predicted by SIQHDy (black dots) for the expectation values of the accessible observables $XI$ (a), $YI$ (b), and $ZI$ (c).
            The dashed vertical line indicates $t=0.1$, up to which 
            the data are used for learning. 
            The learned quantum circuit is recursively applied starting from the known initial state at $t = 0$, and the resulting predicted dynamics are plotted.
            The SIQHDy results are plotted every $10$ sampling points.     
            }
			\label{fig10}
		\end{center}
\end{figure}

Figure~\ref{fig10} compares the true dynamics of the original quantum system in Eq.~(\ref{eq:qsys}) with those predicted by the learned quantum circuit. 
Starting from the known initial state at $t=0$, we recursively apply the learned quantum circuit to predict the dynamics at subsequent sampling times. 
These results demonstrate that the learned quantum circuit accurately reproduces the original dynamics of the expectation values of the accessible observables $XI$, $YI$, and $ZI$ even for $t > 0.1$, after the learning interval.

\subsubsection{Network structure identification of five-spin systems}

Next, we consider a five-spin network system characterized by the Hamiltonian
$$
H = \frac{3}{2} XXIII+IXXII+\frac{3}{2}IIXXI+IIIXX.
$$
We assume that only the single-node observables for the second and fourth spins are accessible
$$M' = \{ IXIII, IYIII, IZIII, IIIXI, IIIYI, IIIZI  \}$$
and measure their expectation values at the sampling times.

Under limited measurement access, identifying the true Hamiltonian structure among all possible candidates is challenging due to the restricted availability of information. Accordingly, in this example, we assume a priori knowledge that the interactions are of the $XX$ type and focus on identifying the underlying network structure.
\begin{figure} [!t]
	\begin{center}
    \includegraphics[width=0.65\hsize,clip]{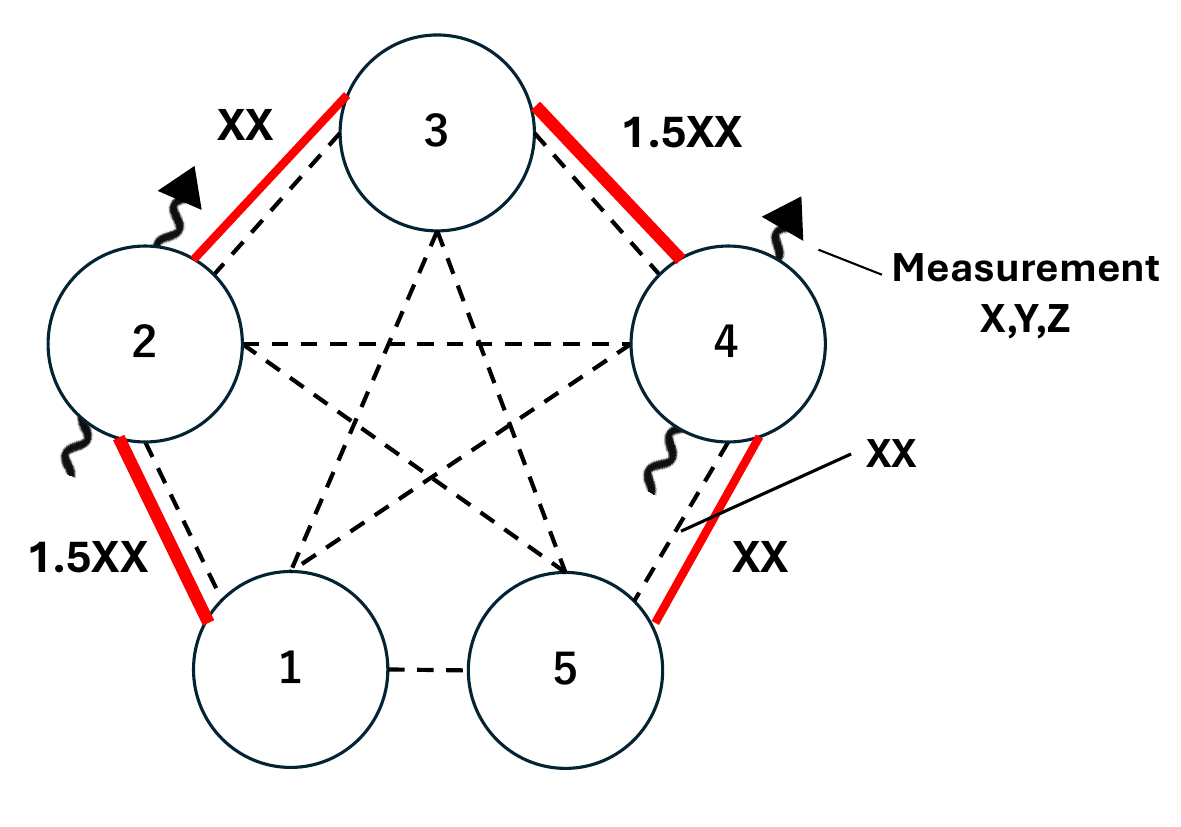}
    \caption{
    Schematic diagram of the five-spin network-structure identification problem considered in this study.
    A priori knowledge of 
$XX$-type interactions is assumed, and all ${}_5C_2 = 10$ possible pairwise couplings are included as candidates (black dashed lines).
The true network couplings (red solid lines) are estimated from the accessible single-spin observables \(X\), \(Y\), and \(Z\) measured on the second and fourth spins.
}
    \label{fig11}
	\end{center}
\end{figure}

Therefore, we restrict the candidate basis quantum circuits to $XX$-type pairwise couplings
and prepare all $ {}_5 C_2 = 10$
possible rotation gates, 
$$
XXIII, XIXII, XIIXI, XIIIX, IXXII,
IXIXI, IXIIX, IIXXI, IIXIX, IIIXX,
$$
and focus on identifying the network structure of the five-spin systems. 
The schematic diagram of the problem setting is shown in Fig.~\ref{fig11}.

In this case, the threshold parameter for the sparsity-promoting procedure in the extended SIQHDy algorithm is set to $\lambda = 0.35 \cdot (2\Delta t)$. As a result, the optimized circuit parameters with nonzero values are obtained as
\begin{align}
\bm{\theta}_{\mathrm{opt}}=
[\theta_{x_{1}x_{2}}, \theta_{x_{2}x_{3}},
 \theta_{x_{3}x_{4}}, \theta_{x_{4}x_{5}}]
= [1.500, 1.000, 1.500, 1.000] \cdot (2\Delta t),
\end{align}
corresponding to 
rotation gates $XXIII, IXXII, IIXXI$, and $IIIXX$, respectively.
These estimated parameters yield a cost function value close to zero, namely 
$8.231 \times 10^{-13}$, confirming the accuracy of the estimation. 

\begin{figure} [!t]
	\begin{center}
			\includegraphics[width=1\hsize,clip]{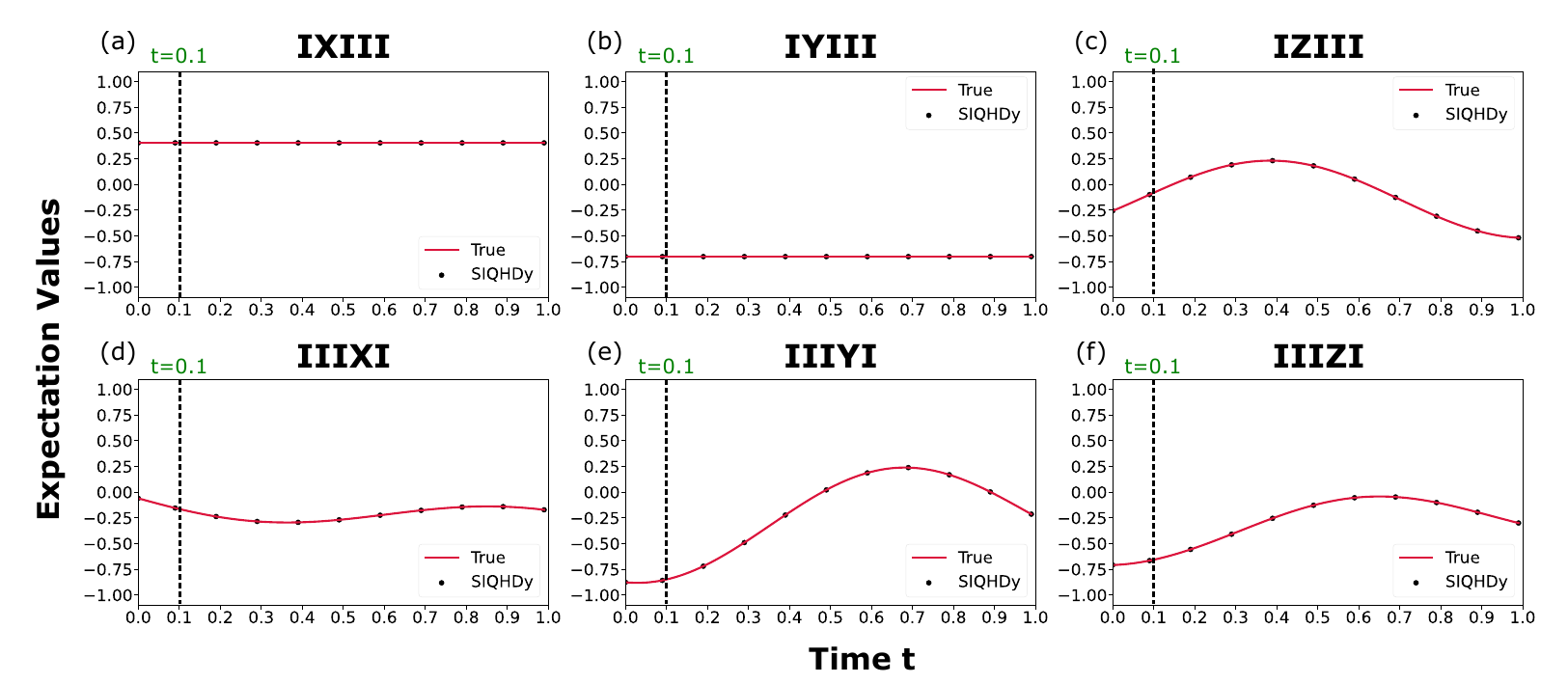}
			\caption{
            Results of the five-spin case.
            (a-f) Comparison between the true dynamics (red solid lines) of the original quantum system and the dynamics predicted by SIQHDy (black dots) for the expectation values of the accessible observables $IXIII$ (a), $IYIII$ (b), $IZIII$ (c), $IIIXI$ (d), $IIIYI$ (e), and $IIIZI$ (f).
            The dashed vertical line indicates $t=0.1$, up to which 
            the data are used for learning. 
            The learned quantum circuit is recursively applied starting from the known initial state at $t = 0$, and the resulting predicted dynamics are plotted.
            The SIQHDy results are plotted every $10$ sampling points.        
            }
			\label{fig12}
		\end{center}
\end{figure}

Figure~\ref{fig12} compares the true dynamics of the original quantum system in Eq.~(\ref{eq:qsys}) with those produced by the learned quantum circuit, in a manner similar to the previous case. It can be seen that the learned quantum circuit accurately predicts the original dynamics of the expectation values of observables $IXIII, IYIII$, $IZIII$, 
$IIIXI, IIIYI$, and $IIIZI$.

\section{Conclusion}

In this study, we introduced sparse identification of quantum Hamiltonian dynamics (SIQHDy), a SINDy-inspired quantum circuit learning framework for estimating quantum Hamiltonian dynamics from time-series data. Through numerical simulations, we demonstrated that SIQHDy can sparsely estimate quantum Hamiltonian dynamics and extract the dominant model structures in single-, three-, and five-spin systems. Additionally, we showed that incorporating sparse optimization enhances robustness to measurement noise, as illustrated in the three-spin case.
Moreover, we proposed an extension of the SIQHDy framework to scenarios with limited accessible observables and numerically demonstrated that the extended SIQHDy method can successfully identify  Hamiltonian dynamics of two-spin systems and reconstruct the network structures of five-spin systems, even under limited measurement access.

Although the present study focuses on quantum Hamiltonian systems, an important direction for future work is to extend the proposed method for the identification of  open quantum systems. Such an extension would be analogous to the ERA-based quantum Hamiltonian identification framework~\cite{zhang2014quantum}, which has been naturally generalized to identify open quantum systems~\cite{zhang2015identification}.

Given its robustness to measurement noise and its applicability under limited measurement access, a promising future direction is to validate SIQHDy using experimental data from real quantum hardware, analogous to the application of the ERA method to nuclear magnetic resonance systems~\cite{hou2017experimental}.
Since practical quantum platforms are inevitably affected by noise and hardware constraints, applying SIQHDy to experimental data would provide valuable insights into its practical applicability and limitations.

In classical nonlinear dynamics, data-driven methods for the analysis and control of complex nonlinear systems have been extensively developed~\cite{brunton2019data}. Extending such approaches to quantum system identification, as demonstrated in this study through the SINDy framework, has the potential to open new directions for analyzing and controlling complex quantum dynamics in many-body systems. 
Moreover, our framework provides a new data-driven paradigm for quantum system identification based on hybrid quantum-classical computation, enabling the direct discovery of interpretable and simple models from time-series data, inspired by data-driven approaches developed for classical complex nonlinear dynamics.

\section*{Acknowledgments}
We acknowledge JSPS KAKENHI JP22K14274 and JST PRESTO JPMJPR24K3 for financial support.

\section*{Appendix}

\subsection{Practical numerical implementation of sparse optimization}
In this appendix, we describe the numerical implementation of the sparse optimization procedure employed in our algorithm.
In the original SINDy framework, sparsity is typically promoted using the sequential thresholded least-squares (STLSQ) algorithm, in which least-squares regression and thresholding are performed alternately and iteratively~\cite{brunton2016discovering}.
However, unlike least-squares optimization, whose optimal solution can be obtained analytically, applying a similar approach in combination with quantum circuit learning becomes computationally expensive, as it requires repeated optimizations of the circuit parameters.
To address this issue, we adopt a simpler and more computationally efficient sparse optimization procedure, as described below.

Instead of directly minimizing the cost function
with the $\ell_{1}$-regularization term
$$
L(\bm{y}, \bm{y}^{out}(\bm{x}, \bm{\theta}))
	=  \left\| \bm{y} - \bm{y}^{out}(\bm{x}, \bm{\theta})  \right\|_{2}^{2} +  \lambda \left\| \bm{\theta} \right\|_{1}.
$$
we perform sparse optimization using the following two-step procedure:
\begin{itemize} 
\item (S1)
The parameters $\bm{\theta}$ are updated by minimizing the cost function without the $\ell_{1}$-regularization term.
\item (S2)
Elements of the parameters $\bm{\theta}$ whose absolute values are smaller than a prescribed threshold $\lambda$ are set to zero.
\end{itemize}
Based on the assumption that the interaction Hamiltonian of the quantum system is sparse, we initialize all parameters as $\theta_i = 0$ for all $i$ and iteratively perform the above two-step procedure.

An illustrative example of the algorithm is provided below, with the threshold value set to $\lambda = 0.25 \cdots 2\Delta t$.
\begin{align*}
&\bm{\theta} = [0.00,0.00,0.00, 0.00]\times 2\Delta t\\
&\overset{1(S1)}{\to}
 [0.71, -0.15,0.33, 0.16]\times 2\Delta t
\overset{1(S2)}{\to}
 [0.71,0, 0.33,0 ]\times 2\Delta t \\
&\overset{2(S1)}{\to}
 [1.00,0, 0.21,0]\times 2\Delta t
\overset{2(S2)}{\to}
 [1.00, 0, 0, 0]\times 2\Delta t
\end{align*}
As illustrated above, basis quantum circuits with parameters whose absolute values fall below the threshold $\lambda$ are gradually removed during the optimization process, resulting in a sparse model structure. 
This thresholding step effectively removes redundant small parameters and retains only the basis quantum circuits that capture the dominant model structures of the underlying dynamics.

In the original SINDy algorithm, the thresholding parameter used to identify the most parsimonious model can be selected based on information criteria for model selection~\cite{mangan2017model}. A similar approach can, in principle, be applied to determine the thresholding parameter in the SIQHDy framework.
%


\end{document}